\documentclass[11pt, a4paper, reqno]{amsart}
\usepackage[centertags]{amsmath}
\usepackage{amssymb}
\usepackage{amsthm}
\usepackage{latexsym}
\usepackage{bbm} 

\numberwithin{equation}{section}
\newcommand{\beq}{\begin{equation}}
\newcommand{\eeq}{\end{equation}}
\newcommand{\tr}{{\rm \, tr\,}}
\newcommand{\al}{\alpha}

\newcommand{\ga}{\gamma}

\newcommand{\la}{\lambda}

\newcommand{\ep}{\epsilon}

\newcommand{\ol}[1]{\overline{#1}}
\newcommand{\ul}[1]{\underline{#1}}
\def\d{\partial}

\def\vi{\varphi}

\def\dl{\dh^\leftrightarrow}
\def\dh{\mathop{\vphantom{\odot}\hbox{$\partial$}}}
\def\dl{\dh^\leftrightarrow}

\def\diag{{\rm diag}}
\newcommand{\mx}{m_X^2}
\newcommand{\my}{m_Y^2}
\newcommand{\mz}{m_Z^2}
\newcommand{\mw}{m_W^2}
\allowdisplaybreaks[4] 
\begin{document}

\title{Violation of Quantum Gauge Invariance in Georgi-Glashow $SU(5)$}
\thanks{Work supported by Swiss National Science Foundation.}
\author[M. Ambauen \& G. Scharf] {Martin Ambauen and G\"unter Scharf \\
\\
\\
Institut f\"ur Theoretische Physik\\
der Universit\"at Z\"urich \\
Winterthurerstrasse 190  \\
CH-8057 Z\"urich Switzerland
\\
\\
{\tt ambauen@physik.unizh.ch}\\
{\tt scharf@physik.unizh.ch}\\
\\
\\}

\begin{abstract}
We check whether the $SU(5)$ model, originally suggested by Georgi and Glashow, is compatible with quantum gauge invariance in first and second order for massive asymptotic gauge fields. We see that this is not the case: the $SU(5)$ grand unified model does not meet with our restrictions from second order gauge invariance. 
$\phantom{}$\\
$\phantom{}$\\
$\phantom{}$\\
{\sc Preprint:} ZH-TH 14/2004  
$\phantom{}$\\
$\phantom{}$\\
{\sc Keywords:} Quantum gauge invariance, Georgi-Glashow Model, Grand Unification  
$\phantom{}$\\
$\phantom{}$\\
{\bf PACS.} 11.15-q Gauge field theories, 12.10-Dm Unified theories and models of strong and electroweak interactions 
12.10-Kt Unification of couplings; mass relations
\end{abstract}
 
\maketitle 
\thispagestyle{empty}
\newpage
\section{Introduction}
In a recent monograph \cite{geist} and related papers quantum gauge theories have been extensively studied in the causal approach.
By means of perturbative quantum gauge invariance [we speak of {\it gauge invariance} in the following] the Weinberg Salam Model of electroweak interaction is deduced by the first three orders of gauge invariance assuming three massive and one massless gauge field. There are some astonishing features in this approach among them the derivation of the Higgs potential in third order without using semi-classical arguments. But the discussion of gauge invariance is even more general. In second order one finds for a very general massive gauge theory restrictions on the masses of the gauge fields and the structure constants of the theory. That means one can explicitly check whether other gauge theories than the Standard Model or theories beyond the Standard Model are gauge invariant.  In the literature grand unified models are normally introduced in a heuristic way.  One considers a specific gauge group and looks on its (irreducible) representations to see whether the particle spectrum fits. Many grand unification schemes are related to the old $SU(5)$ model as it was originally proposed by Georgi and Glashow in 1974 \cite{Georgi:sy}. Therefore it is advisable to check first whether the $SU(5)$ model is gauge invariant at all. 

Our paper is organised as follows: Section \ref{general}  is a brief summary of quantum gauge invariance where the most important restrictions for a general massive gauge theory are given. These restrictions will later be analysed in the case of $SU(5)$. Section \ref{su5} is a review of the Georgi-Glashow model. Most important for our examination is the assignment of the masses and charges to the gauge bosons in the theory. Having this knowledge in mind, we consider in sect.~\ref{Threee}
 which couplings and thus which structure constants can be different from zero. This can in principle be done in two ways: by means of gauge invariance or by charge and colour conservation. We will use both of them here. In section \ref{examination} we examine all gauge restrictions with the mass degeneracy of the Georgi-Glashow model and the structure constants that are not zero. This leads to an equation set that  is very restrictive. We show that the Georgi-Glashow model is not compatible with all of these restrictions. In the following subsections we alter the mass degeneracy for the two new gauge bosons $X$ and $Y$. But this does not help. Again we are led to no go results. The reason for this irritating result is the following: 
Quantum gauge invariance gives many restrictions on masses and coupling constants; hence, if one chooses them according to other heuristic principles, gauge invariance may be spoiled. 

\section{General massive gauge theories}\label{general}
We consider massive gauge theories working with the massive asymptotic fields from the very beginning. We do not use the Higgs mechanism. In our approach we focus on the adiabatically switched $S$-matrix 
\begin{equation}\label{S-Matrix}
S(g)=\mathbbm{1}+\sum_{n=1}^\infty \frac{1}{n!}\int d^4x_1\cdots d^4x_n T_n(x_1,\ldots, x_n)
g(x_1)\cdots g(x_n), 
\end{equation}
and we want to formulate gauge invariance directly for the chronological products $T_n$ which are expressed by asymptotic free fields. This is indeed possible for massless and massive theories \cite{geist} with the definition of a gauge variation $d_Q$ such that
\beq\label{gauge}
\begin{split}
d_QT_n =[Q, T_n] &= d_Q{\rm T}\bigl[T_1(x_1)\cdots T_1(x_n)\bigr]\\
&=: i\sum_{l=1}^n\frac{\d}{\d x_l^\mu}T_{n/l}^\mu(x_1,\ldots,x_l,\ldots, x_n), 
\end{split}
\eeq
where $Q$ stands for the nilpotent gauge charge and $T_{n/l}^\mu$ are time ordered products with a $Q$-vertex $T^\mu_{1/1}$ at $x_l$ \cite{geist}. 

It turns out that our requirement (\ref{gauge}) of gauge invariance is a very strong one. Starting with a general ansatz for the gauge couplings, gauge invariance only, leads to the physically relevant theories. For massless gauge fields it follows that the couplings are essentially of Yang-Mills type which means up to divergence and coboundary terms \cite{Aste:1998iw}, \cite{Dutsch:dp}. Such terms lead to equivalent\footnote{We call two $S$-matrices equivalent if all matrix elements between physical states agree in the adiabatic limit:
$$
\lim_{g \rightarrow 1}(\Phi, {\sf P}S(g){\sf P}\Psi)=
\lim_{g \rightarrow 1}(\Phi, {\sf P}S'(g){\sf P}\Psi),
$$
where ${\sf P}$ is a projection operator on the physical subspace ${\mathcal{F}_{\rm phys}}$ and $\Phi$ and $\Psi$ are arbitrary states in Fock space. Even stronger and more adequate for the massless case is the perturbative version
$$
{\sf P}T_n{\sf P}={\sf P}T_n'{\sf P}+{\rm div}.
$$
Here $T_n$ and $T_n'$ are $n$-point functions corresponding to $S(g)$ and $S'(g)$ respectively. The first definition in terms of the adiabatic limit is implied by this second stronger perturbative one, if the adiabatic limit exists.}
$S$-matrices in the end. In massive gauge theories gauge invariance necessitates the introduction of additional physical scalar fields (Higgs fields) \cite{Dutsch:pf}, \cite{geist}. We have therefore the following field content:

We take $r$ massive and $s$ massless gauge fields $A_a^\mu$,
$a=1,\ldots ,r+s$ together with $(r+s)$ fermionic ghost and anti-ghost
fields $u_a, \tilde u_a$\footnote{The field $\tilde{u}$ is not the adjoint field of $u$; this would be in conflict with the connection between spin and statistics, cf. \cite{geist}, sect.~1.2.}. The masses of a gauge field and the 
corresponding ghost and anti-ghost fields must be equal. We have $m_a=0$ for
$a>r$.
In order to get a gauge charge $Q$, which is nilpotent $Q^2=0,$ as in the massless case,
we have to introduce for every massive
gauge vector field $A_a^\mu(x), a\le r,$ a scalar partner $\Phi_a(x)$ with the
same mass $m_a$, leading to
$$
Q=\int d^3x\,(\d_\nu A^\nu+m\Phi){\dl}_0u,
$$ as discussed in \cite{geist}, sect.~1.5. 
Note, that the scalar and ghost fields appearing in $Q$ are all unphysical
because their excitations do not belong to the physical subspace $\mathcal{F}_{\rm phys}$ and as in the massless case we can express this subspace as 
$${\mathcal F}_{\rm phys}={\rm Ker} Q/{\rm Ran} Q$$
or in any of its equivalent expressions (cf. \cite{geist}, sect.~1.4). For this reason it is convenient to call $u, \tilde{u}$ fermionic ghosts and $\Phi$ bosonic ghosts. 
The additional physical scalar fields (Higgs fields) we denote as $\vi_p, p=1,\ldots ,t$ with arbitrary masses $\mu_p$. We use indices $p,q,\ldots=1,\ldots, t$
from the end of the alphabet to number the Higgs fields, letters
$h,j,k,l,\ldots=1,\ldots, r$ from the middle denote the other massive
scalar fields and $a,b,c,d,e,f,\ldots=1,\ldots, r+s$ is used for the
gauge fields and ghosts.\footnote{The fields are  hermitian. The corresponding mass matrices are already diagonal. This means one will in the case of the electroweak theory directly see the $Z$ and $A^\mu$ field instead of the $W^3$ and the $B$-field (or $W^0$ field \cite{Huang}, (6.25)).}

With those fields the following trilinear
couplings have to be considered:
\[
T_1(x)=T_1^0+T_1^1+\ldots +T_1^{11},
\]
where
\begin{align}
T_1^0&=igf_{abc}(A_{\mu a}A_{\nu b}\d^\nu A_c^\mu-A_{\mu a}u_b\d^\mu
\tilde u_c)\label{T0}\\ 
T_1^1&=ig\frac{m_j^2+m_h^2-m_a^2}{4m_hm_j}f_{ahj}A_a^\mu(\Phi_h\d_\mu\Phi_j-\Phi_j\d_\mu\Phi_h),\\
T_1^2&=ig\frac{m_b^2-m_a^2}{2m_h}f_{abh}A_{\mu a}A^\mu_b\Phi_h\\
T_1^3&=ig\frac{m_j^2-m_h^2+m_a^2}{2 m_j}f_{abh}\tilde u_au_b\Phi_h\label{T3}\\ 
T_1^4&=igf^4_{hjk}\Phi_h\Phi_j\Phi_k =0.\label{T4}
\end{align}
$T_1$ must be skew-adjoint, the Yang-Mills coupling constants $f_{abc}$ are real and
totally antisymmetric, as a consequence of gauge invariance, cf. \cite{geist}, eq.~(3.2.28). The coupling constants in (1.4-7) have been restricted by first order gauge invariance. We have left out the double dots here (and will do so in the following), yet all these products of field operators are normally ordered products of free fields.

As mentioned we must introduce real (Higgs) couplings. Without them we would not achieve second order gauge invariance. We do this by a replacement of the scalar ghosts by Higgs fields:\footnote{Note that this is not the most general form for the couplings. E.~g. in $T_1^6$ the symmetric combination
\beq\label{freedomm}
A_a^\mu(\vi_p\d_\mu\vi_q+\vi_q\d_\mu\vi_p)
\eeq
is not listed because it can be expressed by a divergence. Such terms do not change the physical $S$-Matrix. The same is true for coboundary couplings. By not writing such terms we concentrate on the physically relevant basis of our gauge theories. Our basis is insofar fixed that we can only rotate the generators with mass degeneracy. Such a rotation is done when one combines $W^1$ and $W^2$ to the charged $W^\pm$. And of course also generators for other massive fields with the same mass can be rotated into one another.} 
\begin{align}
T_1^5&=igf^5_{ahp}A_a^\mu(\Phi_h\d_\mu\vi_p-\vi_p\d_\mu\Phi_h)\\
T_1^6&=ig f^6_{apq}A_a^\mu(\vi_p\d_\mu\vi_q-\vi_q\d_\mu\vi_p)\\
T_1^7&=ig m_b f^5_{abp}A_{\mu a}A_b^\mu\vi_p\\
T_1^8&=-ig m_b f^5_{abp}\tilde u_au_b\vi_p\\
T_1^9&=-ig \frac{\mu_p^2}{2m_a}f^5_{hjp}\Phi_h\Phi_j\vi_p\\
T_1^{10}&=ig\frac{\mu_q^2-\mu_p^2}{m_a^2} f^5_{hpq}\Phi_h\vi_p\vi_q\\
T_1^{11}&=igf^{11}_{pqu}\vi_p\vi_q\vi_u.
\end{align}
We have written $T_1$ using all results from first order gauge invariance, where one can relate all coupling constants in (1.4-7) to the pure Yang-Mills couplings $f_{abc}$. For the Higgs couplings only  $f^{5}$,  $f^{6}$ and $f^{11}$ remain unfixed; they are further constrained in second and third order gauge invariance \cite{geist}. But gauge invariance not only fixes all the the structure constants $f^1,\ldots,f^{11} (t=1)$, one can still find more relations between the masses of the gauge bosons and the Yang-Mills structure constants. These relations will be the basis for the forthcoming analysis.

We review here those of the restrictions in second order gauge invariance that are most needed later on. From the tree graphs with external legs $uA\Phi\Phi$ comes a restriction that will be mainly used examining the Georgi-Glashow model. This important condition reads as follows \cite{geist}, (4.4.36)
\beq\label{36}
\begin{split}
\sum_{p=1}^t \Bigl(f^5_{ajp}f^5_{dhp}-f^5_{ahp}f^5_{djp}\Bigr)=&\sum_{c=1}^{r+s}\frac{m^2_j+m^2_h-m^2_c}{2m_hm_j}
f_{dac}f_{chj}\\
&-\sum_{k=1}^r\frac{m^2_k+m^2_j-m^2_a}{m_jm_k}\frac{m^2_k+m^2_h-m^2_d}{4
m_hm_k}f_{ajk}f_{dhk}\\
&+\sum_{k=1}^r\frac{m^2_k+m^2_h-m^2_a}{m_hm_k}\frac{m^2_k+m^2_j-m^2_d}{ 4
m_jm_k}f_{ahk}f_{djk},  \quad (j \ne h).
\end{split}
\eeq
It will be crucial for our analysis that this relation (\ref{36}) is symmetric under interchange of $a$, $d$ and in $h$, $j$ respectively. Interchanging two of these indices gives only a global sign change due to the antisymmetry of the structure constants: Indeed the first summand gets a sign change by interchanging $d$ and $a$ and/or interchanging $h$ and $j$. The second summand is equal minus the third after this interchange, leading to a total sign change only. This comes also true in case $f^5$ is not zero, since also the left hand side in $f^5$ possesses the same 
In the special case with $a=j$ and $d=h$ ($j\ne h$) (\ref{36}) leads to 
\beq\label{37}
\begin{split}
\sum_{p=1}^t f^5_{jjp}f^5_{jjp}=&\frac{1}{2m_h^2}\biggl\{\sum_{c=1}
^{r+s}(m_j^2+m_h^2-m_c^2)f_{jhc}f_{jhc}\\
&-\sum_{k=1}^r\frac{m_k^4-(m_j^2-m_h^2)^2}{2m_k^2}\,f_{jhk}f_{jhk}
\biggr\}.
\end{split}
\eeq
This relation shows, that $f^5_{jjp}$ must be different from zero, because the r.h.s. is not zero in general.
From the sector $uAA\Phi$ one has again a restriction similar to (\ref{36}). We will mainly use the special cases \cite{geist}, (4.4.40)
\beq\label{40}
\begin{split}
m_am_b\sum_{p=1}^t f^5_{aap}f^5_{bbp}\delta_{ac}=&-m_c^2\sum_{d>r}f_{abd} 
f_{bcd}
+\sum_{j=1}^r\frac{f_{abj}f_{bcj}}{4m_j^2}\Bigl[(m_j^2-m_b^2)\times\\ 
&\times(3m_j^2-m_a^2+m_b^2)-m_c^2(m_j^2+m_a^2-m_b^2)\Bigr] \quad (b\ne c) 
\end{split}
\eeq
and  \cite{geist}, (4.4.41)
\beq\label{41}
(m_a^2-m_b^2)\sum_{d>r}(f_{abd})^2=0.
\eeq
For the other restrictions in this and other sectors we will refer to \cite{geist}, sect.~4.4, whenever we will make explicit use of them.
\section{$\mathfrak{su}(5)$ --- a review}\label{su5}
Why do we still work with $\mathfrak{su}(5)$? Experimentally one already knows that $\mathfrak{su}(5)$ cannot be the ultimate unification algebra. For physical theories based on it predict a proton lifetime of \cite{Langacker:1980js}\footnote{From the beginning though there have been attempts to prolong proton lifetime by enlarging $m_{X,Y}$ in order to save $\mathfrak{su}(5)$ \cite{Buras:1977yy} (note added in proof). At the beginning such was giving leeway to $\mathfrak{su}(5)$, but the better and better experimental value for $\sin^2\theta_W(m_X)$ was restraining it soon.  Therefore despite all those modifications in the literature it is today widely believed that $\mathfrak{su}(5)$ is physically definitely ruled out. Mathematically this is not so and explains why many extensions of the model are still discussed \cite{Albright:2000yw}, \cite{Langacker:1994vf}.}
\beq
\tau(p)\approx \frac{m_X^4}{g^2 m_p^5} = 10^{27} \div 10^{31}\;\; {\rm years},
\eeq
($m_X$: unification mass scale, $g$: $\mathfrak{su}(5)$ coupling constant, $m_p$: proton mass) whereas experimental data (Particle Data Group, 1996) predicts for the dominant decay channel in the $\mathfrak{su}(5)$ model
\beq
\tau(p\longrightarrow \pi^0 e^+)\ge 10^{31} - 5\times 10^{32} \;\; {\rm years}.
\eeq
Nevertheless the $SU(5)$ group is believed to be the ingredient of most grand unification groups (as $SO(10)$, $E_6$, $E_8$ and higher $SU(n)$ groups), since it is the smallest simple group that contains the Standard Model (\cite{Chaichian:mm}, ch.~14) and has, as the only rank $n=5$ candidate, except for $SU(3)\times SU(3)$, a complex representation which is necessary for the fermions.  Therefore the  $\mathfrak{su}(5)$ algebra should be compatible with our gauge conditions if such grand unification schemes make physical sense. On the other hand we are interested in the internal consistency of the  $\mathfrak{su}(5)$ gauge theory, not in its relevance in phenomenology.

We start with an introduction to $\mathfrak{su}(5)$ (via \cite{Antoniadis:1981gh}), herewith we are able to check this frame. In the earliest model two representations of $\mathfrak{su}(5)$ are considered: the \ul{24} representation for the super-strong breaking and the \ul{5} representation. They yield a Higgs potential depending on the Higgs multiplets $\Phi$ (from \ul{24}) and $H$ (from \ul{5}) and cross terms of these two fields \cite{Sherry:1979sz}
\beq\label{hleft}
\begin{split}
V(H, \Phi)=&-\frac{1}{2}\mu^2\tr \Phi^2 +\frac{1}{4}a(\tr \Phi^2)^2+\frac{1}{2}b\, \tr \Phi^4-\frac{1}{2}\nu^2H^\dagger H+\frac{1}{4}\lambda (H^\dagger H)^2\\
&+\alpha H^\dagger H \tr \Phi^2+\beta H^\dagger \Phi^2 H,
\end{split}
\eeq
where the terms with $\alpha$ and $\beta$ describe those cross terms; without them the minimisation is done in \cite{Buras:1977yy}, giving rise to a minimal model. For mixed minimisation see \cite{Sherry:1979sz}, where the $\mathfrak{su}(5)$ Higgs potential is extensively studied. In the case $\alpha\ne 0$ and $\beta \ne 0$ the vacuum expectation values take on the form
\beq\label{Higgseps}
\left<H\right>= h\left(
\begin{array}{c}
0 \\ 0 \\ 0 \\ 0 \\ \frac{1}{\sqrt{2}}, 
\end{array} \right),
\;\left<\Phi\right>=\nu \;{\rm diag}(1, 1, 1,-\frac{3}{2} -\frac{1}{2}\epsilon, -\frac{3}{2} +\frac{1}{2}\epsilon),\;\, \epsilon=\frac{3}{20}
\frac{\beta}{b}\frac{h^2}{\nu^2} + O\bigl(\frac{h^4}{\nu^4}\bigr).
\eeq
The parameter $\epsilon$ is a combination of the parameters in (\ref{hleft}) that comes out from minimisation of this Higgs potential. If $\epsilon$ is zero one speaks of {\it minimal $\mathfrak{su}(5)$}.

\subsection{Normalisation according to Georgi-Glashow}\label{GG}
We outline here the $\mathfrak{su}(5)$ as it was proposed by Georgi and Glashow in \cite{Georgi:sy} and described at length by Langacker in his famous review article \cite{Langacker:1980js}, sect.~3.3. which --- in particular for this section --- originated in \cite{Buras:1977yy}. The gauge fields are written as a sum over the generalised Gell-Mann matrices $\la^i$  as they are listed in \cite{Hayashi:1976rs} in the case of $\mathfrak{su}(5)$ and one gets
\beq\label{Bosonmatrix}
A=\sum_{i=1}^{24}\frac{A^i\la^i}{\sqrt{2}}
=\left( 
\begin{array}{ccccc}
G^1_1-\frac{2B}{\sqrt{30}} & G_2^1 & G^1_3 & \ol{X}^1 & \ol{Y}^1  \\
G_1^2 & G^2_2-\frac{2B}{\sqrt{30}} & G_3^2 & \ol{X}^2 & \ol{Y}^2  \\
G_1^3 & G_2^3 & G^3_3-\frac{2B}{\sqrt{30}} & \ol{X}^3 & \ol{Y}^3  \\
X_1 & X_2 & X_3 & \frac{W^3}{\sqrt{2}}+\frac{3B}{\sqrt{30}} & W^+ \\
Y_1 & Y_2 & Y_3 & W^- & -\frac{W^3}{\sqrt{2}}+\frac{3B}{\sqrt{30}} \\
\end{array} \right),
\eeq 
\beq
\tr A =0
\eeq
with
\beq
\begin{split}
G_1^1&=G^3/\sqrt{2}+G^8/\sqrt{6}\\
G_2^2&=-G^3/\sqrt{2}+G^8/\sqrt{6}\\
G_3^3&=-2G^8/\sqrt{6},
\end{split}
\eeq
The normalisation depends on the representation one works with: In the trace condition \cite{Weinberg:kr}, (21.5.1)
\beq
\tr \{T_\al T_\beta\}=N_D\delta_{\al\beta}
\eeq
$N_D$ can take any value. For the fundamental representation in $\mathfrak{su}(n)$  its value is $N_D =2$  \cite{Duetsch:1994ur}. The factor
\beq
1/\sqrt{2n(n+1)}= 1/\sqrt{2}\sqrt{30}
\eeq
is a special normalisation due to \cite{Georgi:jb}, eq.~(13.4). 
The normalisation in the literature defers sometimes from this specific one.\footnote{Cf. \cite{O'Raifeartaigh:vq}, \cite{Huang}, \cite{Langacker:1980js} (for physical normalisations), \cite{Langacker:1980js} and \cite{Georgi:jb} (for general $\mathfrak{su}(n)$ traceless generators). Especially the $B$ and $W^3$ fields are differently normalised.} Only the trace condition is demanded, since the fields must be hermitian. The off-diagonal $G$'s are combined like in the  $\mathfrak{su}(3)$ case as
\beq
\begin{split}
G_2^1&=G^1/\sqrt{2}+iG^2/\sqrt{2}\\
G_1^2&=G^1/\sqrt{2}-iG^2/\sqrt{2}
\end{split}
\eeq
and for the electroweak part
\beq\label{Wplusminus}
W^\pm=(W^1\pm iW^2)/\sqrt{2}
\eeq
and the same is done with the super-heavy particles $X$ and $Y$ in this model
\beq\label{XYplusminus}
\ol{X}^1=(X^1+iX^2)/\sqrt{2}, \; 
X^1=(X^1-iX^2)/\sqrt{2} \; \text{etc.}
\eeq
For the off-diagonal entries in $A$ (\ref{Bosonmatrix}) there is no freedom for rotation; they correspond directly to the hermitian generators for the gauge bosons. Those generators can be rotated however if they are part of the same multiplet with mass degeneracy. This freedom allows for the rotation (\ref{Wplusminus}) and  (\ref{XYplusminus}).
The diagonal generators can be rotated as this is done in the electroweak case in order to obtain the neutral $Z$ boson and the photon $A^\mu$ (\cite{Huang}, (6.25)) from the diagonal  $W^3$ and $B$ generator with the Weinberg angle $\theta_W$
\beq\label{ZGamma}
\begin{split}
Z_\mu&=\cos\theta_W W^3_\mu-\sin\theta_W B_\mu\\
A_\mu&=\sin\theta_W W^3_\mu+\cos\theta_W B_\mu.
\end{split}
\eeq
In $\mathfrak{su}(5)$ we have 
four diagonal generators: two correspond to the diagonal gluon generators and the other two look like
\beq
\begin{split}
W_3&=\diag (0,0,0,1/\sqrt{2},-1/\sqrt{2})\\
B & = \diag\frac{1}{\sqrt{30}} (2, 2, 2,-3,-3).
\end{split}
\eeq
We denote them here as in the electroweak case, because the same rotation (\ref{ZGamma}) is carried out leading to the generators for the $Z$ and the photon $A^\mu$ (we denote it as $\ga$ for short)
\beq
Z=\left( 
\begin{array}{ccccc}
+\frac{2}{\sqrt{30}}\cos\theta_W &  &  &  &  \\
 &  \hspace{-1cm}+\frac{2}{\sqrt{30}}\cos\theta_W &  & \hspace{1.5cm}\mbox{\huge 0}  &  \\
 &  & \hspace{-1cm}+\frac{2}{\sqrt{30}}\cos\theta_W &  &  \\
\mbox{\huge 0} &  &  &  \hspace{-1.4cm}-\frac{1}{\sqrt{2}}\sin\theta_W - \frac{3}{\sqrt{30}}\cos\theta_W &  \\
 &  &  &  &  \hspace{-1.9cm}\frac{1}{\sqrt{2}}\sin\theta_W - \frac{3}{\sqrt{30}}\cos\theta_W \\ 
\end{array} \right),
\eeq
\beq
\ga=\left( 
\begin{array}{ccccc}
-\frac{2}{\sqrt{30}}\sin\theta_W &  &  &  &  \\
 &  \hspace{-1cm}-\frac{2}{\sqrt{30}}\sin\theta_W &  &  \hspace{1.5cm}\mbox{\huge 0} &  \\
 &  & \hspace{-1cm} -\frac{2}{\sqrt{30}}\sin\theta_W &  &  \\
\mbox{\huge 0} &  &  &  \hspace{-1.3cm}\frac{1}{\sqrt{2}}\cos\theta_W + \frac{3}{\sqrt{30}}\sin\theta_W &  \\
 &  &  &  &  \hspace{-1.8cm}-\frac{1}{\sqrt{2}}\cos\theta_W + \frac{3}{\sqrt{30}}\sin\theta_W \\ 
\end{array} \right). 
\eeq
Based on these generators one can calculate the masses for the gauge bosons.
\subsection{Mass relations in $\mathfrak{su}(5)$}\label{Massinsu5}
It is possible to express the gauge boson masses in terms of the parameters of the Higgs potential and the coupling constants therein. For the mass relations we first consider the case when in (\ref{Higgseps}) $\ep = 0$ (minimal $\mathfrak{su}(5)$) and $h=0$ for simplification. For getting the mass terms one starts with the expression for the covariant derivative for $\Phi$ which is in the adjoint representation \cite{Cheng and Li}, eq.~(14.30)
\beq
\begin{split}
D_\mu \Phi&= \d_\mu \Phi + ig[A_\mu, \Phi]\\
       &=D_\mu \Phi' + ig[A_\mu, \left<\Phi\right>].
\end{split}
\eeq  
$A_\mu$ denotes here the matrix (\ref{Bosonmatrix}) and $\Phi '= \Phi -\hspace{-0.1cm}\left<\Phi\right>$ is the shifted Higgs field in matrix form \cite{Cheng and Li}, eq.~(14.29).  As in the Standard Model one  interprets $|D_\mu \Phi|^2$ as a term for kinetic energy. Therein the scalar  $g^2|[A_\mu, \left<\Phi\right>|^2$ is the desired mass term. Looking for expressions for the masses of the super-heavy particles $X$ and $Y$ we have to project out the relevant contributions from the $A$-matrix (\ref{Bosonmatrix}). For the first $X$-particle these are the generators
\beq
\lambda_a=\left( 
\begin{array}{ccccc}
0 & 0 & 0 & 1 & 0  \\
0 & 0 & 0 & 0 & 0  \\
0 & 0 & 0 & 0 & 0  \\
1 & 0 & 0 & 0 & 0  \\
0 & 0 & 0 & 0 & 0  
\end{array} \right), \;
\lambda_b=\left( 
\begin{array}{ccccc}
0 & 0 & 0 & -i & 0  \\
0 & 0 & 0 & 0 & 0  \\
0 & 0 & 0 & 0 & 0  \\
i & 0 & 0 & 0 & 0  \\
0 & 0 & 0 & 0 & 0  
\end{array} \right). 
\eeq
We then calculate the commutators\footnote{A faster way (\cite{Cheng and Li}, Problems and Solutions) is to write the commutators like
\begin{displaymath}
[A_\mu,\left<\Phi\right>]^j_k=(A_\mu)^j_k(\Phi_k-\Phi_j),
\end{displaymath}
where
\begin{displaymath}
\left<\Phi\right>_k^j=\Phi_k\delta^j_k.
\end{displaymath}
Here one sees directly that if $\Phi_k=\Phi_j$ then the gauge field is massless.}
\beq
[\la_a, \left<\Phi\right>]=-[\la_b,\left<\Phi\right>]=\left( 
\begin{array}{ccccc}
0 & 0 & 0 & -2.5 & 0  \\
0 & 0 & 0 & 0 & 0  \\
0 & 0 & 0 & 0 & 0  \\
+2.5 & 0 & 0 & 0 & 0  \\
0 & 0 & 0 & 0 & 0  
\end{array} \right).
\eeq
Taking the trace of the square, this gives the value for the mass of the $X$-bosons
\beq
m_X^2=\frac{25}{8}g^2\nu^2.
\eeq
One gets the same value for the mass of the $Y$-bosons, when $h=0$ as it was assumed so far. In the more general case with two Higgs-multiplets we have the mass formula \cite{Buras:1977yy}
\beq
\mathcal{M}_{ab}=g^2[-\frac{1}{8}\tr\{[\la^a,\left<\Phi\right>][\la^b,\left<\Phi\right>]\}+\frac{1}{8}\left<H\right>^\dagger\{\la_a,\la_b\}\left<H\right>];
\eeq
for $m_X$ this does so far not change anything, but to $m_Y$ will be added a term from $H$. If we also take cross terms of the two Higgs multiplets into account i.e. $\ep\neq 0$  in (\ref{Higgseps}) we get for both masses an $\ep$ correction. Finally we get the mass relations, as one can find them in the literature \cite{Chaichian:mm},  \cite{Antoniadis:1981gh} up to normalisation:
\begin{enumerate}
\item $m_W^2=\frac{1}{8}g^2 h^2+\frac{1}{4}\epsilon^2 g^2 \nu^2$
\item $m_Z^2=\frac{8}{15}g^2 h^2$
\item $m_X^2=\frac{1}{8}(5+\epsilon)^2 g^2 \nu^2$
\item $m_Y^2=\frac{1}{8}(5-\epsilon)^2 g^2 \nu^2 + \frac{1}{4}g^2 h^2$
\item eight massless gluons
\item one massless photon.
\end{enumerate}

\subsection{Charges of the $\mathfrak{su}(5)$ bosons}\label{Chargeinsu5}
Apart from the mass relations for the gauge bosons we will need information about the structure constants. One can of course write a program to obtain the structure constants for the normalisation we work with. Yet there is freedom and one is not sure whether exactly these values for the structure constants come out from gauge invariance. In order to circumvent this problem we will adopt a more general treatment for the structure constants in the following, based on charge conservation and gauge invariance itself. Therefore we have a glance at charges in  $\mathfrak{su}(5)$ in this section.
As for smaller  $\mathfrak{su}(n)$ algebras one can calculate the charge operator $Q$ as a superposition of the diagonal generators of the algebra. In the fundamental representation $Q$ looks like \cite{Cheng and Li}, (14.12)
\beq
Q=\diag (-1/3, -1/3, -1/3, 1, 0).
\eeq
The charges of the  particles in the $A$-matrix $A_{ij}$ are $Q(A_{ij})=Q_i-Q_j$, where $Q_i, \, Q_j$ are the diagonal entries in $Q$. Gluons and $W$'s remain neutral/charged as they are in the Standard Model, the new $X$ and $Y$ particles carry electric charge: 
\beq
Q_X=-1/3-1=-4/3, \;\;\; Q_Y=-1/3-0 =-1/3.
\eeq
The charges are different from $\pm 1$ and each multiplet has different values. This is helpful when considering the coupling structure for the model. With charge conservation one can look for possible couplings between the gauge bosons which will reduce the number of parameters in our second order relations considerably.

\subsection{Notation}
Before looking at the coupling structure, let us agree upon a little remake in the notation: Instead of $\ol{X}^1, \ol{X}^2,  \ol{X}^3$ and $X^1, X^2, X^3$ we denote the $X$ particle multiplet just with the numbers $i=1,\ldots,6$ and similarly the $Y$ particles with $j=1,\ldots,6$ if only one of the two multiplets is involved. In case both are involved we denote them by $X^i, \; i=1,\ldots,6$ and $Y^j, \; j=1,\ldots,6$ respectively. This notation is more general; the notation in (\ref{Bosonmatrix}) could be misleading. 

\section{Coupling structure in Georgi-Glashow  $\mathfrak{su}(5)$}\label{Threee}
\subsection{Possible couplings by virtue of charge conservation}\label{charge}
We have studied in the last sections the Georgi-Glashow model in order to be sure what  masses  and charges the bosons in this special $\mathfrak{su}(5)$ scheme have. As we noticed we will here not work with specific values for the structure constants as one could calculate them easily by commuting the generalised Gell-Mann matrices. When doing this one finds that our second order relations cannot be fulfilled. Yet, as noticed, the freedom for the diagonal generators could be a stumbling block and one cannot be conclusive at all. We would have to allow for rotations which rakes us even more unknown parameters! It is therefore advisable to let the structure constants be as general as possible. Here they ought to be antisymmetric and real only.

In order to discuss charge conservation we suppose the coupling structure of the charged super-heavy fields as $X^+=X^1+iX^2$ and $X^-=X^1-iX^2$, $W^\pm=W^1\pm iW^2$ which is mirrored by  $A$ (\ref{Bosonmatrix}) in the Georgi-Glashow model. From charge conversation we know which couplings among them are possible. What can we say then on the coupling structure among the hermitian fields as $W^1$,   $W^2$,  $X^1$,  $X^2$ etc.?

From charge conservation we know e.g. that
$f_{X^+ X^- W^\pm} = 0$; this implies $f_{X^1 X^2 W^2} $ $= 0$ and $f_{X^1 X^2 W^1} = 0$, because
\beq
\begin{split}
f_{X^+ X^- W^+}=& \, f_{X^1 X^1 W^1}+if_{X^1 X^1 W^2}+if_{X^2 X^1 W^1}-f_{X^2 X^1 W^2}\\
&-if_{X^1 X^2 W^1}+f_{X^1 X^2 W^2}-f_{X^2 X^2 W^1}+if_{X^2 X^2 W^2}\\
=&-2if_{X^1 X^2 W^1}+2f_{X^1 X^2 W^2}       
\end{split}
\eeq
due to antisymmetry of the $f$'s. This shows that $W^{1,2}$ cannot couple to two $X$ bosons if they are charged in the way it was described in subsection \ref{Chargeinsu5}. The same holds for the couplings of the form $f_{X^+ X^- Y^\pm}$. Hence $f_{X^1 X^2 Y^i}=0,\; i=1,\ldots,6$. The only left massive boson is then the $Z^0$. Let us see whether this can couple to two $X$. We have  
$f_{X^+ X^- Z^0}\ne 0$ and then 
\beq
\begin{split}
f_{X^+ X^- Z}&=f_{X^1 X^1 Z}-if_{X^1 X^2 Z}+if_{X^2 X^1 Z}+f_{X^2 X^2 Z}\\
&=-2if_{X^1 X^2 Z}.       
\end{split}
\eeq
So if as assumed $f_{X^+ X^- Z}$ is not zero we have also $f_{X^1 X^2 Z^0}\ne 0$. 
The arguments above hold in the case of two $Y$'s as well of course, only the amount of charge is  different. From charge conservation, couplings of the form  $f_{X^+ Y^- W^\pm}$  are possible, $f_{X Y Z}$ not. But here the couplings for the hermitian generators cannot be coercively ruled out. We will show this using gauge invariance.  Similar arguments though can be used for the couplings of the form $f_{X X X}$ or $f_{Y Y Y}$. 

\subsection{Additional constraints from gauge invariance}\label{gaugeandacharge}
In this subsection we will show how far gauge invariance can restrict the coupling structure in $\mathfrak{su}(5)$. 
We start with couplings of the form  $f_{X W Z}$. We go back to (\ref{40}) and analyse it for $a=X^2$, $b=W^1$ and $c=W^2$. When the masses $m_X$ and $m_W$ are different we have no couplings $f_{X W d}$, with $d>r$ due to eq.~(\ref{41}): 
\beq 
\begin{split}
0=& +\sum_j f_{X^1 W^1 j}f_{W^1 W^2 j}[(m_j^2-m_W^2)(3m_j^2-m_X^2+m_W^2)\\
     & -m_W^2(m_j^2+m_X^2-m_W^2)]\\
  = &\, f_{X^1 W^1 Z}f_{W^1 W^2 Z}[3m_Z^4-m_Z^2m_X^2-3m_Z^2m_W^2],
\end{split}
\eeq
where the second equality holds because we know from the electroweak relation (cf. below \ref{weaktwo}), that $W^1$ and $W^2$ couple only to the $Z$. Similar arguments hold for $f_{Y W Z}$, $f_{X W W}$, $f_{X Z  Z}$ and the like. 
We conclude the discussion of purely massive couplings and show that  $f_{X Y Z}=0$. We go back to (\ref{40}) and analyse it for $a=X^2$, $b=X^1$ and $c=Y^1$:
\beq 
\begin{split}
0=&(3m_W^2-3m_X^2-m_Y^2) [f_{X^1 X^2 W^1}f_{X^1 Y^1 W^1}+f_{X^1 X^2 W^2}f_{X^1 Y^1 W^2} ]\\
&+(3m_Z-3m_X^2-m_Y^2)f_{X^1 X^2 Z}f_{X^1 Y^1 Z}\\
&-m_Y^2 \sum_i f_{X^1 X^2 X^i}f_{X^1 Y^1 X^i} \\
&+(2m_Y^2-3m_X^2)\sum_j f_{X^1 X^2 Y^j}f_{X^1 Y^1 Y^j},       
\end{split}
\eeq
where we did not write couplings to massless indices here because of (\ref{41}) again.
One can repeat the same choice for all combinations of the particles of the two multiplets being left with an equation set that restricts the (sums of) structure constants. But this is not conclusive, since it could be possible that the structure constants adopt any value, not just in the range of $\pm 2$ because in a relation as e.g. in the electroweak case 
\beq
\Bigl(\frac{f_{W^1 W^2 \ga}}{f_{W^1 W^2 Z}}\Bigr)^2=\tan^2\theta
\eeq
the mixing angle is not necessarily given by the value $\sin^2\theta=3/8$, but can even lie close to $\pi/2$; $\tan^2\theta$ can then be a huge number. This means that the structure constants could be of values near to the mass ratios $m_{X}/m_Z$, $m_{Y}/m_Z$ etc. 
In a future publication we will analyse such equation systems in detail. One can check any mass degeneracy and is not forced to use arguments based on charge conservation as one does not know in general how these new gauge particles are charged. For the Georgi-Glashow model though we can take  the results from section \ref{charge} into account and are left with the relation
\beq 
0= f_{X^1 X^2 Z}f_{X^1 Y^1 Z}(3m_Z-3m_X^2-m_Y^2)
\eeq
since in the other products of structure constants always one factor is zero. We end up  with a mass relation that is not true for the Georgi-Glashow model and conclude that  $f_{X^1 Y^1 Z}$ is zero, since  $f_{X^1 X^2 Z}$ is not.

So far we have considered purely massive couplings. For couplings between massive and massless bosons gauge invariance alone does constrain the coupling structure. In this case we do not need charge conservation. For example going to eq.~(\cite{geist}, (4.4.43)), setting $a=h=X^1$ (index 1) and $b=\gamma$ we note that the left hand side is zero because $f^5_{\gamma\gamma p}$ is zero, and we see that 
\beq\label{lambdawe}
\begin{split}
0=& - \sum_{d=1}^r (f_{1\gamma d })^2 \frac{(m_d^2+m_X^2)^2}{2m_X^2m_d^2} + 2\sum_{d=1}^r (f_{\gamma 1 d})^2 \frac{m_d^2}{m_X^2}\\
=& \sum_{d=1}^r (f_{1\gamma d })^2 \Bigl[2\frac{m_d^2}{m_X^2}- \frac{(m_d^2+m_X^2)^2}{2m_X^2m_d^2}\Bigr]
\end{split}
\eeq
if we now allow the massive index $d$ to run through $X^i,\; i=1,\ldots, 6$ only, this relation is best fulfilled:
\beq
0= \Bigl[2\frac{m_X^2}{m_X^2}-\frac{4 m_X^4}{2 m_X^4}\Bigr] \sum_{i=1}^6 (f_{1 \gamma X^i})^2.
\eeq
If the sum over $d$ ran over other massive indices ($d=Z^0$, $W^{1,2}$, $Y^j$, $j=1,\ldots,6$) beside the $X$ bosons, the mass factors would not compensate. Thus these structure constants vanish. 
The same holds true if we replace the photon index $\gamma$ with a gluon index here, i.e. couplings that are forbidden by ``color conservation'' can also be ruled out by gauge invariance: we show here that couplings of the form $f_{\la\la Z}$ can be suspended. This can best be demonstrated using eq.~(\ref{40}) with the content $a=c=\la_1$, $b=\la_2$, which leads to
\beq
0=0-\frac{3}{4}f_{\la^1\la^2 Z}^2m_Z^2.
\eeq
Since $m_Z$ is not zero and because we can choose here any two gluon indices  we have reached our aim, i.e. $f_{\la\la Z}=0$.
Also the type $f_{W W \la}$ is easily shown to vanish: in (\ref{36}) we put
$d=\la^1$, $a=\la^2$, $h=W^1$ and $j=W^2$ and see
\beq
0=\sum_{i=1}^8 f_{\la^1\la^2\la^i} f_{\la^i W^1 W^2}\cdot 1 -0 +0.
\eeq
And because the gluons do self-couple and since we can here go through all gluon indices we have necessarily $f_{\la W W}=0$.
As seen in (\ref{41}) all couplings for two massless to one massive gauge boson do vanish generally. The same is true for the coupling of one massless boson to two bosons of different non-vanishing masses, as for $f_{\la X Y}$.
We summarise our possible non-vanishing couplings in table 1. In the sequence of this work we will look for relations among these couplings and the masses involved. We start with the electroweak relation and look whether it still holds with this coupling structure in the bigger model.
\begin{table}[h]
\begin{tabular}{c|c}
{\sc Couplings of massless gauge bosons} & {\sc Purely massive couplings}\\
\hline
                                         &  \\ 
                 $f_{\la \la \la}$        & $f_{W W Z}$ \\
                 $f_{Y Y \la}$           & $f_{X X Z}$ \\
                 $f_{X X \la}$           & $f_{Y Y Z}$ \\
                 $f_{Y Y \ga}$          & $f_{X Y W}$ \\
                 $f_{X X \ga}$          &  \\
                 $f_{W W \ga}$        &  \\
                                        
\vspace{0.2cm}                                           
\end{tabular}
\caption{Possible non-vanishing structure constants. The same index always means a different particle of the same multiplet, e.g. $W W$ corresponds to $W^1 W^2$ and $\la \la \la$ means $\la_i \la_j \la_k$ for $i < j < k\in \{1,\ldots,8\}$.}  
\end{table}

\subsection{Electroweak relation}\label{weaktwo}
In  $\mathfrak{su}(5)$ the electroweak relation (\cite{geist}, (4.6.5)) should still hold true, also if for the Weinberg angle a very different value can result. In this subsection we dwell briefly on this point and show that the electroweak relation indeed formally remains true, underlining that the couplings make sense. In \ref{gaugeandacharge} we did already use the fact that $W^1$ and $W^2$ couple only to the $Z$. One can see here the consequences if we had not assumed this. The electroweak relation would be changed and we would not see the subalgebra $\mathfrak{su}(2)\oplus \mathfrak{u}(1)$ anymore. Actually we should see again that
\beq\label{elecweak}
\frac{f_{W^1 W^2 Z}^2}{f_{W^1 W^2 \gamma}^2}=\frac{m_Z^2}{m_W^2}-1
\eeq
holds. For this we go to (\ref{37}) with $j=W^1$, $h=W^2$ and equate it with the same equation but setting $h=Z$, which is always possible with the same index $j$ here,
\beq
\begin{split}
&\frac{m_{W_{1}}^2+{m_{W_{2}}^2}}{2m_{W_{2}}^2} \biggl( \sum_{c>r} f_{W^1 W^2 c}^2 + f_{W^1 W^2 Z}^2\biggr)  \\
&-\frac{m_Z^2}{2m_W^2}f_{W^1 W^2 Z}
-\frac{m_Z^4-m_{W_{1}}^4+m_{W_{2}}^4-2m_{W_{1}}^2m_{W_{2}}^2}{4m_Z^2}f_{W^1 W^2 Z}\\
\stackrel{!}{=}&\frac{1}{2}f_{W^1 Z W^2}^2-\frac{1}{2m_Z^2}\frac{m_{W_{2}}^4-m_{W_{1}}^4+m_Z^4-2m_{W_{1}}^2m_Z^2}{2m_W^2}f_{W^1 Z W^2}.
\end{split}
\eeq
and we arrive again at
\beq\label{elmoreg}
\frac{\sum_{c>r}f_{W^1W^2c}^2}{f_{W^1W^2 Z}^2}=\frac{4m_{W_{2}}^2(m_{W_{1}}^2-m_{W_{2}}^2)+2m_Z^2(m_Z^2-m_{W_{2}}^2)}{m_Z^2(m_{W_{1}}^2+m_{W_{2}}^2)};
\eeq
we can here put $m_{W_{1}}=m_{W_{2}}$ (cf. \cite{geist}, (4.3.19)). 
We see that only electroweak particles are involved here and find again the electroweak relation, as claimed, making apparent, that we have indeed included $\mathfrak{su}(2)\oplus \mathfrak{u}(1)\subset\mathfrak{su}(5)$. And as there are no restrictions for the gluons ($\mathfrak{su}(3)$) from gauge invariance: we have even the inclusion $\mathfrak{su}(3) \oplus \mathfrak{su}(2)\oplus \mathfrak{u}(1)\subset\mathfrak{su}(5)$.
This shows that our coupling structure makes sense in the electroweak case. Had we allowed couplings of the form $ f_{W^1W^2 X}$ or  $ f_{W^1W^2 Y}$ we would have a much more complicated relation than (\ref{elecweak}).

\section{Explicit check of $\mathfrak{su}(5)$}\label{examination}
We are ready now for the examination of the Georgi-Glashow model. Let us summarise the field content we work with below
\begin{enumerate}
\item 8 massless gluons
\item 1 massless photon
\item 2 massive charged $W$ bosons with mass $m_W$ 
\item 1 massive uncharged $Z^0$ with mass $m_Z$
\item 6 massive charged $X$ bosons with masses $m_X$
\item 6 massive charged $Y$ bosons with masses $m_Y$.
\end{enumerate}
For the structure constants we look at table 1. All of the second order relations have to be fulfilled simultaneously. We start with the purely massive couplings, since gauge invariance then is very restrictive. This is especially true when one inserts couplings involving gauge bosons with four different masses. How many equations do we have in this case?
In the products of structure constants recurrences are not allowed which reduces the 256 possibilities to 4! combinations. From these 24, not all are different from each other because of symmetries in the indices of some second order relations. For (\ref{36}) and eq.~(\cite{geist}, (4.4.51)) this means that only $\left( \begin{array}{c} 4 \\ 2\end{array} \right)= 6$ choices remain. They all yield non-trivial results. In eq.~(\cite{geist}, (4.4.39)) and in eq.~(\cite{geist}, (4.4.42))\footnote{In the last column there is a misprint: instead of $+2f_{bcj}f_{ahj}$ one should read $-2f_{bcj}f_{ahj}$, therefore this relation is not different from (4.4.39). Another sign should be adjusted in (4.4.51): on the left hand side there is a total sign change.}  12 cases remain, because they are only symmetric in two indices. The remaining relations cannot be analysed here, because there are only three or two indices to be inserted. Those relations with one or more identical indices in the products of structure constants are not accessible to the case where one treats four different gauge bosons.

We will basically work with (\ref{36}) in this paper since it is already conclusive  (as mentioned above in the purely massive case this relation complies with eq.~(\cite{geist}, (4.4.51) so this relation is confirmed from two independent sectors). 
Let us compile below, what we have found:
(\ref{36}), multiplied by $2m_hm_j$, setting $d=X^1$, $a=W^1$, $h=Z$, $j=Y^1$ we get:
\beq\label{36erste}
\begin{split}
0 =& \sum_{j=1}^6 f_{X^1Y^jW^1}f_{Y^1 Y^jZ}\\
   &+\frac{m_Y^2+m_X^2-m_W^2}{2m_X^2}\sum_{i=1}^6 f_{X^iY^1W^1}f_{X^1 X^iZ}\\
   &+\frac{m_Y^2+m_W^2-m_X^2}{2m_W^2} f_{W^1W^2Z}f_{X^1Y^1W^2},
\end{split}
\eeq
where we have cancelled a factor $m_Z^2$. We go on with (\ref{36}) with $d=W^1$, $a=Y^1$, $h=X^1$, $j=Z$ which yields:
\beq\label{36zweite}
\begin{split}
0 =& \sum_{i=1}^6 f_{X^iY^1W^1}f_{X^1 X^i Z}\\
   & +\frac{m_Y^2+m_X^2-m_W^2}{2m_Y^2}\sum_{j=1}^6 f_{Y^1 Y^jZ}f_{X^1 Y^jW^1}\\
   & +\frac{m_W^2+m_X^2-m_Y^2}{2m_W^2} f_{X^1Y^1W^2}f_{W^1 W^2Z}
\end{split}
\eeq
and with $d=X^1$, $a=Y^1$, $h=Z$, $j=W^1$:
\beq\label{36dritte}
\begin{split}
0 =& f_{X^1Y^1W^2}f_{W^1 W^2Z}\\
   & + \frac{m_X^2+m_W^2-m_Y^2}{2m_X^2}\sum_{i=1}^6 f_{X^iY^1 W^1}f_{X^1 X^iZ} \\
   & + \frac{m_Y^2+m_W^2-m_X^2}{2m_Y^2}\sum_{j=1}^6 f_{Y^1 Y^jZ}f_{X^1 Y^j W^1}
\end{split}
\eeq
which is the last equation that involves only the masses $m_X$, $m_Y$ and $m_W$. The remaining relations are
(\ref{36}) setting $d=Z$, $a=W^1$, $h=X^1$, $j=Y^1$: 
\beq\label{36ersteZ}
\begin{split}
0 =& f_{W^1W^2Z}f_{X^1 Y^1 W^2}\\
   & +\frac{2m_X^2-m_Z^2}{2m_X^2}\sum_{i=1}^6 f_{X^iY^1W^1}f_{X^1 X^i Z}\\
   & +\frac{2m_Y^2-m_Z^2}{2m_Y^2}\sum_{j=1}^6 f_{X^1Y^jW^1}f_{Y^1 Y^jZ},  
\end{split}
\eeq
where a common factor $m_Y^2+m_X^2-m_W^2$ has cancelled,
(\ref{36}) with $d=Z$, $a=Y^1$, $h=X^1$, $j=W^1$:
\beq\label{36zweiteZ}
\begin{split}
0 =& -\sum_{j=1}^6 f_{Y^1 Y^jZ}f_{X^1Y^j W^1}\\
   & -\frac{2m_X^2-m_Z^2}{2m_X^2}\sum_{i=1}^6 f_{X^i Y^1 W^1}f_{X^1 X^iZ} \\
   & -\frac{2m_W^2-m_Z^2}{2m_W^2} f_{X^1Y^1W^2}f_{W^1W^2Z}
\end{split}
\eeq
and (\ref{36}) with $d=Z$, $a=X^1$, $h=W^1$, $j=Y^1$:
\beq\label{36dritteZ}
\begin{split}
0 =
&\sum_{i=1}^6 f_{X^1 X^iZ}f_{X^i Y^1W^1}\\
&+\frac{2m_W^2-m_Z^2}{2m_W^2}f_{X^1 Y^1 W^2}f_{W^1 W^2 Z}\\
   & +\frac{2m_Y^2-m_Z^2}{2m_Y^2} \sum_{j=1}^6 f_{X^1 Y^jW^1}f_{Y^1Y^jZ}.
\end{split}
\eeq
We see that we have many more relations than structure constants, since the (sums of) structure constants are the same in every restriction. Hence it is advisable to eliminate the three (sums of) structure constants
\beq\nonumber
\begin{split}
A&:=\sum_i f_{X^1 X^i Z}f_{X^i Y^1 W^1}\\
B&:=\sum_j f_{Y^1Y^jZ}f_{X^1 Y^jW^1}\\
C&:=f_{X^1Y^1W^2}f_{W^1W^2Z} 
\end{split}
\eeq
in the following equation set
\beq\left|
\begin{array}{l}\label{set}
0 = 2 B \mx \mw + A (\my + \mx - \mw)\mw + C(\my + \mw - \mx)\mx \\
0 = 2 A \my \mw + B (\my + \mx - \mw) \mw + C (\mw + \mx - \my) \my \\
0 = 2 C \mx \my + A (\mx + \mw - \my) \my + B (\my + \mw - \mx)\mx \\
0 = 2 C  \my  \mx + A (2\mx - \mz)\my + B (2 \my - \mz)\mx \\
0 = 2 B \mx \mw + A (2\mx - \mz)\mw + C (2\mw - \mz)\mx \\
0 = 2 A \my \mw + C (2\mw - \mz)\my + B(2\my - \mz)\mw 
\end{array}\right| .
\eeq
All equations are linear in $A, B$ and $C$. Therefore an elimination with a computer program like Mathematica poses no problem. Elimination is, of course, only possible if $A,B$ and $C$ are different from zero. We will assume this at first. If  $C$ was zero we would be left with a trivial theory, where none of the new gauge bosons would couple to the $W^1$ (or by interchange $W^2$). We will later discuss what happens when one or both of the sums  $A$ or  $B$ are zero which could well be. First we do an elimination of $A$ and $B$ using the first three equations of (\ref{set}). From this elimination the restriction
\beq
\begin{split}
& C \mx\my(m_W^6 - m_W^4\mx - \mw m_X^4 + m_X^6 - 
          m_W^4\my + 2\mw\mx\my \\
&- m_X^4\my -\mw m_Y^4 - \mx m_Y^4 + 
          m_Y^6) = 0
\end{split}
\eeq
results with the following solutions:
\beq\label{poss1}
\begin{split}
 C &= 0 \\
 \my &= 0\\
 \mx &=0 \\
 \mw &= \mx + \my \\
 \mw &= \mx -  \my  \\
 \mw &= -\mx +  \my .
\end{split}
\eeq

Now taking the fourth and the fifth equation in (\ref{set}) and eliminating $A$ we get the following simple restriction
\beq
C\my - B\mw = 0 
\eeq
or if $\mw \neq 0$
\beq\label{res1}
B=C\frac{\my}{\mw}.
\eeq
From the last two equations in (\ref{set}) we have the additional restriction 
\beq\label{res2}
C \my \mz (-2\mw + \mz) = B \mw (2 (\mx + \my) - \mz) \mz.
\eeq
Elimination of $B$ in (\ref{res1}) and  (\ref{res2})  yields
\beq
C \mx \my (\mw + \mx + \my - \mz) \mz = 0
\eeq
which would mean
\beq
\mw = \mz -\mx -\my;
\eeq
but there is, compared with (\ref{poss1}), no common solution, we see thus already from this sector alone that with both $A$ and $B$ non zero the Georgi-Glashow model with these specific mass degeneracies and charges for the super-heavy bosons is not possible. 

What happens if we set $A=0$? This has no physical implications because it only means that the {\it sum} vanishes not the individual products of structure constants. We take the fourth and the fifth equation of  (\ref{set}) as before and eliminate $B$ getting
\beq
C \mx (2 (\mw + \my) - \mz) \mz = 0
\eeq
with the only non-trivial solution
\beq \label{\mw}
\mw=\frac{\mz}{2}-\my .
\eeq
Taking the last two equations in (\ref{set}) for elimination of $B$ we get
\beq
C \mx \mz (-2\mw + \mz) = 0
\eeq
with the only non-trivial solution
\beq
\mw = \frac{\mz}{2}
\eeq
and comparing with (\ref{\mw}) this would lead to $\my =0$ which is not true. 

Now we assume $B=0$ and take the fourth and the fifth equation of (\ref{set}) as before and eliminate $A$ getting
\beq
C\mx\my\mz = 0
\eeq
and this leads directly to $C=0$ which would only be true if none of the new bosons coupled to the $W^2$. Finally if $A=0$ {\it and} $B=0$ we are again forced to set $C=0$. 

We can also derive no-go results from other sectors separately and in combination, taking equations from different sectors.
This means we have {\it physically} ruled out the Georgi-Glashow model.  {\it Mathematically} there was not yet a contradiction. We are just forced to set some of the variables to zero which is physically not tenable. Trivial solutions with some of the masses or sums of structure constants set to zero always survive.

{\it Remark:}
Note that (\ref{set}) does also not allow for the Georgi-Glashow $\mathfrak{su}(5)$ model, where $\mx =\my$, which would mean that only one Higgs multiplet was present.  This is not the Georgi-Glashow model in its minimal version but is often used as a first approximation. 

\subsection{Other mass degeneracies, no-go Results}\label{NOGO}
We consider in the following the  $\mathfrak{su}(5)$ model with less mass degeneracy for the super-heavy gauge bosons $X$ and $Y$. The charges for $X$ and $Y$ remain unchanged. This too leads to no-go results. First let us assume that per two $X$ particles the mass is equal, i.e. $m_{X^1}=m_{X^2} \ne m_{X^3}=m_{X^4} \neq  m_{X^5}=m_{X^6}$ and similarly for the $Y$ bosons.

We have thus the following field content:
\begin{enumerate}
\item 8 massless gluons
\item 1 massless photon
\item 2 massive charged $W$ bosons with mass $m_W$ 
\item 1 massive uncharged $Z^0$ with mass $m_Z$
\item 6 massive charged $X$ bosons with masses $m_i,\; i=1,3,5$
\item 6 massive charged $Y$ bosons with masses $m_j,\; j=1,3,5$,
\end{enumerate}
where we write $m_1$ for $m_{X^1}$ and $m_{X^2}$ etc. when there is no confusion, i.e. when not both the $X$ and $Y$ particles are appearing together. It is now to be investigated, whether such a theory is conformable with second order gauge invariance.

Let us first go to (\ref{40}), setting $a=X^1$, $b=X^2$ and $c=X^3$ and with the masses $m_{X^1}=m_{X^2}\ne m_{X^3}$. For the subscript $X^i$ we will write just the index $i$ for short, leading to
\beq\label{40'}
0=-m_3^2\sum_{d>r}f_{12d}f_{23d} + \sum_{j=1}^{r}\frac{1}{4}\bigl[3m_j^2-3m_1^2-m_3^2\bigr]f_{12j}f_{23j}.
\eeq
Note that the first sum over the massless indices is zero due to eq.~(\ref{41}), because its left hand side is not zero in our case
\beq\label{41'}
(m_1^2-m_3^2)\sum_{d>r}(f_{23d})^2=0.
\eeq
Hence $f_{23d}$ must vanish. As seen in section \ref{charge} the only possibility for a massive boson to couple to two $X$ bosons, say $X^1$ and $X^2$, is the $Z^0$ boson. Therefore (\ref{40'}) reduces to

\beq\label{40''}
0=\frac{1}{4}\bigl[3m_Z^2-3m_1^2-m_3^2]f_{12Z}f_{23Z}
\eeq
and in the case that both $f_{12Z}$ and $f_{23Z}$ are not zero we could deduce the relation

\beq\label{40'''}
3m_1^2+m_3^2=3m_Z^2.
\eeq
Such a relation is not tenable for a theory with super-heavy masses $m_1$ and $m_3$. We are therefore required to set one or both structure constants to zero. We first discuss the case when $f_{12Z}\ne 0$ then $f_{23Z}\ne 0$ and in the end $f_{12Z}= 0$ {\it and} $f_{23Z}=0$.

{\bf Case $f_{12Z}\ne 0$}. In this case $f_{23Z}=0$ $(c=3)$ and by setting the index $c=4$ we have the same relation (\ref{40'}), as $m_3=m_4$ leading to $f_{24Z}=0$. In the same line of reasoning we have finally for $c=5,6$
   
\beq\label{fuenf}
3m_1^2+m_5^2=3m_Z^2
\eeq
leading to $f_{25Z}=0$ and $f_{26Z}=0$. Note that $c= b$ is not allowed in this relation and that is why we did not set $c=2$ in this case. 

Now we interchange the index $a$ and $b$ in eq.~(\ref{40}) giving
\beq\label{abe}
0=\frac{1}{4}\bigl[3m_Z^2-3m_3^2-m_1^1]f_{21Z}f_{13Z}
\eeq
with $f_{21Z}= -f_{12Z}\neq 0$, hence $f_{13Z}=0$. We deduce in the same manner: $f_{1cZ}=0\;$ with varying $c$ as above. In the following table we summarise what we have achieved so far; the second column refers to $a$ and $b$ interchanged in eq.~(\ref{40}):
\begin{table}[h]\label{skonstI}
\begin{tabular}{c|c|c}
 $f_{12Z}\ne 0$ & $f_{23Z}=0$ & $f_{13Z}=0$ \\
                & $f_{24Z}=0$ & $f_{14Z}=0$ \\
                & $f_{25Z}=0$ & $f_{15Z}=0$ \\
                & $f_{26Z}=0$ & $f_{16Z}=0$ 
\end{tabular}
\end{table}

The next structure constant with one index equal to 3 that could be different from zero is $f_{34Z}$ and assuming it to be different from zero\footnote{The arguments remain also true if we make an other choice, e.g. if  $f_{35Z}\ne 0$ only $f_{46Z}$ can be different from zero, all others have to vanish.}  we find
\begin{table}[h]\label{skonstII}
\begin{tabular}{c|c|c}
 $f_{34Z}\ne 0$ & $f_{45Z}=0$ & $f_{35Z}=0$ \\
                & $f_{46Z}=0$ & $f_{36Z}=0$ \\
                & $f_{41Z}=0$ & $f_{31Z}=0$ \\
                & $f_{42Z}=0$ & $f_{32Z}=0$ 
\end{tabular}
\end{table}

\hspace{-.4cm}and assuming $f_{56Z} \ne 0$ we have
\begin{table}[h]\label{skonstIII}
\begin{tabular}{c|c|c}
 $f_{56Z}\ne 0$ & $f_{61Z}=0$ & $f_{51Z}=0$ \\
                & $f_{62Z}=0$ & $f_{52Z}=0$ \\
                & $f_{63Z}=0$ & $f_{53Z}=0$ \\
                & $f_{64Z}=0$ & $f_{54Z}=0$
\end{tabular}
\end{table}

\hspace{-.4cm}where the last column corresponds again to $a$ and $b$ interchanged. As a result we have now three structure constants that couple the $Z^0$ particle to two massive $X$'s. We have now to see whether this goes along with our other second order relations. 
Setting $a=1$, $b=3$, $d=2$ and $j=4$ in eq.~(\cite{geist}, (4.4.51)) we have specified this relation in the same way as eq.~(\ref{40}) above with $m_1=m_2\ne m_3=m_4$ but now with only three possible couplings for $Z^0$ to the $X$ bosons
\beq\label{51'}
\begin{split}
0 =& \sum_{c>r}f_{13c}f_{2c4}\frac{m_3^2+m_1^2}{2m_3}  
+ f_{13Z}f_{2Z4}\frac{m_3^2-m_Z^2+m_1^2}{2m_3}  \\
&+ f_{14Z}f_{23Z}\frac{m_Z^2+m_3^2-m_1^2}{4m_3m_Z^2}(m_Z^2-m_3^2+m_1^2) \\
&- f_{34Z}f_{21Z}\frac{m_Z^2+m_3^2-m_3^2}{4m_3m_Z^2}(m_Z^2-m_1^2+m_1^2) \\
=&- f_{34Z}f_{21Z}\frac{m_Z^2}{4m_3};
\end{split}
\eeq
where we first note that $f_{13c}=0$ for the massless indices $c$ according to (\ref{41}). Then due to our reasoning above also $f_{13c}=0$ and $f_{14c}=0$ holds for massive indices i.e. the coupling to $Z^0$. The last equality produces again a result that is not tenable for a reasonable gauge theory. One or both of the structure constants must be zero then. If $f_{12Z}$ is not zero we have $f_{34Z}=0$ and with the same reasoning ($b=5, j=6$) $f_{56Z}=0$. If both $f_{12Z}$ and $f_{34Z}$ vanish, we cannot constrain $f_{56Z}$: In any case we thus have only one coupling that survives.

Let us try this last possibility with only one coupling from $Z^0$ to two $X$'s assuming $f_{12Z}$ not to be zero. In (\ref{36}) we find, setting $a=W^1$, $d=W^2$, $h=X^1$ and $j=X^2$, a relation that is not contradictory (including couplings to the $Y$ bosons) but the special case (\ref{37}) gives a further strong constraint demanding a symmetry in the coupling of the different $X$'s: Setting $j=X^1$ (index 1), $h=X^2$ we find with $m_1=m_2$:
\beq\label{37'}
\begin{split}
\sum_{p=1}^{t}f_{11p}^5f_{11p}^5&= \frac{1}{2m_1^2}\biggl[\sum_{c=1}^{r+s}(2m_1^2-m_c^2)f_{12c}f_{12c} - \frac{m_Z^2}{2}f_{12Z}f_{12Z} \biggr] \\
&= 1\sum_{c>r}f_{12c}f_{12c} + (1- \frac{m_Z^2}{2m_1^2}-\frac{m_Z^2}{4m_1^2})f_{12Z}f_{12Z} \\
&=   \sum_{c=1}^{r+s}f_{12c}f_{12c} -\frac{3}{4}\frac{m_Z^2}{m_1^2} f_{12Z}f_{12Z}\\
&\ne 0 
\end{split}
\eeq   
As in the derivation of the electroweak theory we use here again the fact that the left hand side of (\ref{37'}) remains the same while changing the index $h$. Let us set $h=X^3$ with now $m_1\ne m_3$
\beq \label{37''}
\begin{split}
\sum_{p=1}^{t}f_{11p}^5f_{11p}^5&= \frac{1}{2m_1^2}\biggl[\sum_{c=1}^{r+s}(m_1^2+m_3^2-m_c^2)f_{13c}f_{13c} - \frac{m_Z^4 - (m_1^2-m_3^2)^2}{2m_Z^2}f_{13Z}f_{13Z}\biggr] \\
&= 0 
\end{split}
\eeq     
because $f_{13c}$ is zero with $c$ standing for both massive and massless indices. We therefore see that one coupling alone leads to a contradictory result.

{\bf Case $f_{23Z}\ne 0$}.
If $f_{23Z}\ne 0$ we see this time, varying the index $a$ in eq.~(\ref{40}), that all structure constants of the form $f_{a2Z}$ and, by interchanging $b$ and $c$ which is possible because the mass factor does not change, also those of the form  $f_{a3Z}$ must vanish:\begin{table}[h]
\begin{tabular}{c|c|c}
 $f_{23Z}\ne 0$ & $f_{12Z}=0$ & $f_{13Z}=0$ \\
                & $f_{42Z}=0$ & $f_{43Z}=0$ \\
                & $f_{52Z}=0$ & $f_{53Z}=0$ \\
                & $f_{62Z}=0$ & $f_{63Z}=0$ 
\end{tabular}
\end{table}
and assuming $f_{14Z}\ne 0$ we find that all structure constants of the form $f_{4cZ}$ with and $f_{1cZ}$ with $c=2,3,5,6$ must vanish and we have only $f_{56Z}$ as non-vanishing structure constant left. With these three remaining structure constants we go through the same line of reasoning as in the case $f_{12Z}\ne 0$ above.

{\bf Case} $f_{12Z}= 0$ and $f_{23Z}=0$.
If both structure constants vanish, we could start again with say $f_{13Z}$ and $f_{3cZ}$ in the same way as in the other cases. If then one structure constant is chosen to be different from zero only three non-vanishing structure constants survive in the same way as in the other cases or we would  have the trivial solution with all structure constants zero.

\subsection{A last-ditch attempt}
Finally we discuss other possible mass degeneracies for our heavy bosons $X$ and $Y$. First let us assume that per three $X$ particles the mass is equal, i.e. $m_{X^1}=m_{X^2} = m_{X^3} \ne m_{X^4} = m_{X^5}=m_{X^6}$ and similarly for the $Y$ bosons. That means this time the field content is:
\begin{enumerate}
\item 8 massless gluons
\item 1 massless photon
\item 2 massive charged $W$ bosons with mass $m_W$ 
\item 1 massive uncharged $Z^0$ with mass $m_Z$
\item 6 massive charged $X$ bosons with masses $m_i,\; i=1,2$
\item 6 massive charged $Y$ bosons with masses $m_j,\; j=1,2$.
\end{enumerate}
We have now two possibilities in eq.~(\ref{40}), one with only one mass $m_X$ in the mass factor, setting $a=1, \; b=2,\;c=3$ or $a=4, \; b=5,\;c=6$ or one with two different masses in the mass factor $m_{X^1}$ and $m_{X^2}$, setting e.g. $a=1, \; b=2$ but $c=4,5$ or $6$. The case of one common mass we have treated above. We therefore only consider the second alternative with two masses, but
this turns out to lead to the same relations as discussed in section \ref{NOGO} with three masses for the six $X$ or $Y$ bosons. Also here one can reduce the number of non-vanishing structure constants to three. Finally the last step, as in (\ref{37'}), is as well recovered by setting in eq.~(\ref{37}) $j=1$ and $h=2$ and $h=4$ respectively.

\section{Conclusions}
In our analysis we have come to the conclusion that the $SU(5)$ model proposed by Georgi and Glashow is not compatible with perturbative quantum gauge invariance,  a tough result. What could be wrong with this unification scheme?
Specific to the Georgi-Glashow model was the mass degeneracy and the charges of the new super-heavy particles $X$ and $Y$. The rest comes directly with  $SU(5)$ group itself, as the number of generators in its adjoint representation. 
We think that the arguments (in the fundamental representation) for the charges of the diagonal gauge bosons are also a necessity. Therefore only two ways could save $SU(5)$. 

Either the Higgs potential that is introduced by hand and where the mass degeneracy comes from is wrong. One therefore would have to look for other breaking patterns \cite{Sherry:1979sz}. 
Or the hermitian generators for the new gauge bosons apart from the electroweak ones have to be differently combined, as e.g. $X^+=X^1 +i(X^2 +X^3)$ etc. This would take the freedom mentioned after  (\ref{freedomm}) into account.  If this was the case, one would find a solution in second order gauge invariance in a more general approach, i.e. when not restricting the couplings using charge conservation arguments as we did it for the Georgi-Glashow model, since we know there exactly how the gauge bosons arise from their hermitian generators. This is under study; one can then discuss various mass degeneracies again. 
 
Finally there is a third radical possibility, namely that grand unification is not realised at all. The idea of unification was certainly fruitful in physics. But it is not certain that unification must always go to the very end. Indeed, there are arguments against this. The first doubts come from experiment. With present data the running coupling constants of weak, electromagnetic and strong interactions do not meet in a common point. One must introduce supersymmetry to save the situation. A second argument is theoretical. From quantum gauge invariance it follows only that the Yang-Mills coupling constants are structure constants of a {\it reductive} Lie algebra \cite{geist}. These algebras are direct sums of Abelian and simple Lie algebras. Why should nature not make explicit use of an Abelian term $\mathfrak{u}(1)$, if it is mathematically possible? It is an aesthetical prejudice only that simple algebras are nicer than reductive ones. 

\providecommand{\bysame}{\leavevmode\hbox to3em{\hrulefill}\thinspace}

\end{document}